# Longwave-transparent low-emissivity material

**Authors**

Yue Zhang[1,4]†, Longnan Li[1]†*, Junyan Dai[2]†, Xiaowen Zhang[1,4], Qunyan Zhou[2], Naiqin Yi[1,4], Ruizhe Jiang[2], Fei Zhu[1,4], Xiaopeng Li[3], Mengke Sun[2], Jiazheng Wu[1,4], Xinfeng Li[1,4], Xiangtong Kong[1,4], Ziai Liu[1], Yinwei Li[3], Qiang Cheng[2], Yiming Zhu[3], Tie Jun Cui[2]*, Wei Li[1]*

**Affiliations**

[1]GPL Photonics Laboratory, State Key Laboratory of Luminescence Science and Technology, Changchun Institute of Optics, Fine Mechanics and Physics, Chinese Academy of Sciences, Changchun 130033, China.
[2]State Key Laboratory of Millimeter Waves, Southeast University, Nanjing 210096, China.
[3] Terahertz Technology Innovation Research Institute, Terahertz Spectrum and Imaging Technology Cooperative Innovation Center, Shanghai Key Laboratory of Modern Optical System, University of Shanghai for Science and Technology, Shanghai 200093, China.
[4]University of Chinese Academy of Sciences, Beijing 100049, China.
†These authors contributed equally to this work
*Correspondence to weili1@ciomp.ac.cn; tjcui@seu.edu.cn; longnanli@ciomp.ac.cn

**Abstract**

   Low emissivity (low-e) materials are crucial for conserving thermal energy in buildings, cold chain logistics and transportation by minimizing unwanted radiative heat loss or gain. However, their metallic nature intrinsically causes severe longwave attenuation, hindering their broad applications. Here, we introduce, for the first time, an all-dielectric longwave-transparent low-emissivity material (LLM) with ultra-broadband, high transmittance spanning 9 orders of magnitude, from terahertz to kilohertz frequencies. This meter-scale LLM not only achieves energy savings of up to 41.1% over commercial white paint and 10.2% over traditional low-e materials, but also unlocks various fundamentally new capabilities including high-speed wireless communication in energy-efficient buildings, wireless energy transfer with radiative thermal insulation, as well as non-invasive terahertz security screening and radio frequency identification in cold chain logistics. Our approach represents a new photonic solution towards carbon neutrality and smart city development, paving the way for a more sustainable and interconnected future.

**Teaser**

   A scalable all-dielectric low-emissivity longwave-transparent material enables synergetic thermal and information management.

# MAIN TEXT

## Introduction

Suppressing thermal radiation between objects at different temperatures is a fundamental capability for arresting unnecessary energy losses. Uncontrolled radiative heat exchange drives heating and cooling loads, which collectively account for a substantial portion of global heating, ventilation, and air conditioning (HVAC) energy, estimated over 40%(*1*).By selectively inhibiting thermal radiation, engineered materials impose a radiative barrier that curbs this thermodynamic inefficiency(*2, 3*). High-emissivity radiative-cooling materials have been studied extensively for passive cooling(*4-14*), but they increase heating requirements in cold climates. In contrast, low-emissivity (low-e) materials minimize net radiative heat exchange(*15-20*), delivering energy savings over 4 times greater than radiative-cooling approaches(*12, 21*). Beyond building envelopes and cold-chain packaging(*21-27*), low-e coatings play critical roles in spacecraft thermal control, electronics heat management, textile-based personal thermal comfort, and industrial process insulation(*28, 29*)—underscoring their broad applicability for fundamental thermal-photonic regulation. As such, extensive efforts have been devoted to expanding low-e materials, including metals(*20, 24, 30-33*), conductive metal oxides(*27*), silver nanowires(*34*), MXene(*35*), and metal-polymer composites(*36*).

However, all existing low-e materials, regardless of their material composition, fundamentally rely on the infrared reflectivity of metallic components (Fig. 1A-i). Due to the intrinsic Drude dispersion, where permittivity monotonically decreases with frequency(*37*), high infrared reflectivity inherently induces high reflectivity at longer wavelengths (THz to kHz wave, Fig. 1A-i). Such limitation not only fundamentally handicaps the information and energy transfer, but also severely hinders their practical applications (Fig. 1A-i and Fig. 1B). For example, metallic low-e windows(*38-41*), wall envelopes(*38, 42-44*), and roofs(*42, 44*) in buildings drastically attenuate wireless signals (Fig. 1B)(*38, 42*). Furthermore, metal foil low-e barrier packaging - ubiquitous in cold chain logistics - obstructs wireless security screening and tracking, disabling safety inspection against explosives, firearms, drugs and biochemical threats(*28, 45*). As such, achieving low infrared emissivity along with broadband longwave transparency (Fig. 1A-ii and Fig. 1B) remains highly sought after yet inherently unattainable with all existing low-e materials.

Here, we introduce, for the first time, a longwave-transparent low-emissivity material (LLM) that enables synergetic thermal and ultra-broadband information management capabilities (Fig. 1A-ii). This novel photonic material leverages all-dielectric, abundant infrared-transparent microparticles with dimensions comparable to thermal wavelengths, effectively serving as a light-scattering medium (Fig. 1A-ii). It achieves a mid-infrared (MIR, 2.5-14 μm) reflectance of 92.2% and solar reflectance (0.3-2.5 μm) of 94.4%, while maintaining transparency across longwave spectra covering 9 orders of magnitude, from terahertz to kilohertz frequencies, encompassing all wireless communication bands (Fig. 1A-ii and Fig. 1C). These distinctive features of LLM not only reduces heating energy consumption by up to 41.1% compared to commercial white paint and 10.2% compared to low-e film, but also unlocks a wide range of fundamentally new capabilities (Fig. 1C). These includes high-speed wireless communication in energy-efficient buildings, wireless energy transfer with radiative thermal insulation, as well as non-invasive terahertz security screening and radio frequency identification (RFID) in cold chain logistics. Our approach represents a new photonic solution to revolutionize low-e thermal insulation technology, contributing to carbon neutrality and advancing informatization (Fig. 1C)(*46*).

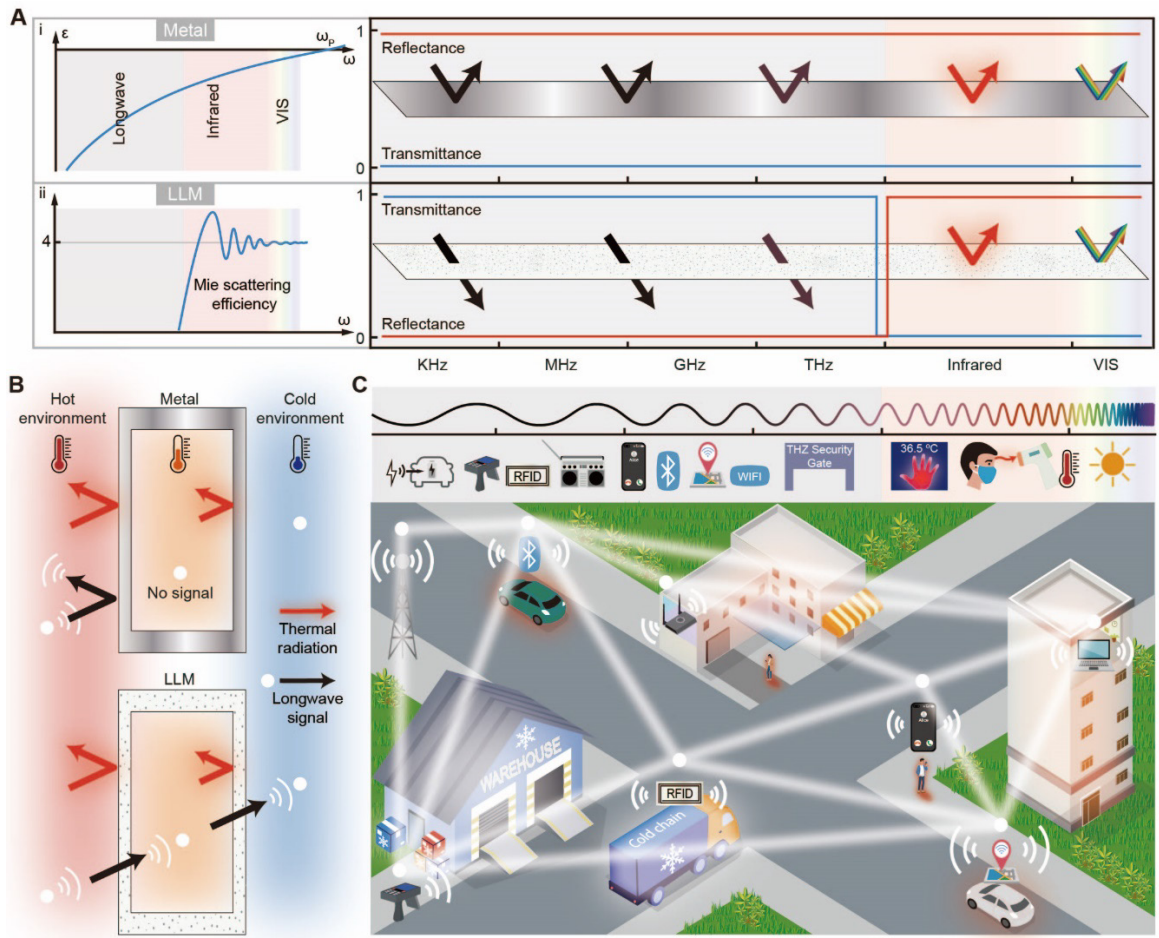

**Fig. 1. Conceptual illustration of LLM.** (**A**) (i) Schematic of (left) Drude dispersion characteristics and (right) the typical photonic properties for metals. Due to Drude dispersion, the permittivity of metals decreases monotonically from high to low frequencies. The high infrared reflectivity inevitably leads to strong attenuation in the longwave range. (ii) (left) Working mechanism and (right) ideal photonic properties of LLM. The low-loss, metal-free microparticles in LLM strongly scatter infrared light while remaining transparent to longwave frequencies. This enables not only high infrared reflectance but also high transmittance across the ultra-broadband longwave range (THz to kHz). (**B**) Schematic comparison of (top) metallic low-e material and (bottom) LLM in radiative thermal insulation. Metallic low-e materials substantially reduce longwave signal strength, while LLM enables high transmission of longwave signals, allowing for synergetic thermal and information management. (**C**) (top) Working frequencies for various applications and (bottom) real-world scenarios (e.g., building insulation, transportation and cold chain logistics) that require both radiative thermal insulation and high longwave transmittance.

## Results

### Concept and realization of LLM

To achieve the desired properties for the LLM, infrared-transparent materials that function as scattering media are essential. Suitable candidates include Barium Fluoride (BaF$_2$), Caesium Iodide (CsI), Magnesium Fluoride (MgF$_2$), Potassium Bromide (KBr), Potassium Chloride (KCl), Sodium Chloride (NaCl), Zinc Selenide (ZnSe), Zinc Sulfide (ZnS), and low-absorptive polymers like polyolefins, including Polyethylene (PE), polypropylene(*47*), and thermoplastic elastomers such as styrene-ethylene-butylene-styrene (SEBS) block copolymer. These materials can act as microparticle scatterers or as matrix materials that form light-scattering air voids, thereby achieving low-e properties through multiple approaches and enhancing the design versatility of the LLM. For our proof-of-concept demonstration, we selected NaCl as light-scattering medium due to its favorable optical properties, availability, and ease of preparation. The lattice vibration frequency of NaCl crystals lies in the far infrared region ($\lambda > 20$ μm), resulting in a near-zero extinction coefficient from visible to MIR wavelengths(*48*) — a key feature for achieving low-e properties in the MIR region through efficient light scattering. We engineered NaCl crystal microparticles (Fig. 2A-i) with sizes ranging from 700 nm to 40 μm (fig. S1 and fig. S2), derived from recycled brine waste sourced from the seawater desalination (Fig. 2A). These microparticles leverage Mie scattering (Fig. 2A-ii, fig. S1, fig. S3 and Supplementary Text 1) to enhance broadband MIR reflectance.

The LLM was fabricated in two forms: a pure LLM sample optimized for optical performance and film-type LLM sample for scalable fabrication. For the pure LLM, NaCl microparticles were cast in a mold and subjected to a high-temperature sintering (fig. S4), achieving 92.2% MIR reflectance and 94.4% solar reflectance (Fig. 2B, inset). For scalable fabrication, NaCl microparticles were blended with SEBS binder (fig. S5) to create a durable framework, then applied to a nanoporous polyethylene (NanoPE) film (fig. S6) using blade coating. After solvent evaporation, the NaCl microparticles were uniformly distributed (Fig. 2A-iii), resulting in a large-scale LLM (2 m × 0.2 m × 850 μm) (Fig.2A-iv) with an average MIR reflectance of 85.2% and solar reflectance of 96.6% (Fig. 2B and fig. S7), effectively blocking solar heating in enclosed spaces(*23, 24, 49*). With a simple encapsulation in NanoPE film, LLM show high stability when it was simulated as a thermal insulation wall envelop (fig. S8, fig. S9 and Supplementary Text 3), critical for real-world applications.

To evaluate the longwave transmission capabilities of the LLM, we conducted measurements across a wide frequency spectrum, from 100 kHz to 1.60 THz. For comparison, we selected commercially available low-e radiant barriers commonly applied in building insulation (insets in Fig. 2C and fig. S10). These metallic low-e materials demonstrated extremely low transmittance (< 1%) across the entire tested longwave spectrum, with aluminum foil, for instance, transmitting only 0.07%, corresponding to a 32.47 dB attenuation in the 1-18 GHz communication band (Fig. 2C, fig. S10, fig. S11 and Supplementary Text 2). In contrast, LLM exhibited significantly higher transmission, averaging over 80% across an ultra-broadband range from kHz to THz. This performance marks a fundamental shift from the conventional approach that is intrinsically narrowband, and often require trade-offs, such as reducing infrared reflectivity to improve microwave transmittance in low-e glasses(*41, 50*). LLM effectively overcomes the inherent trade-off between low-e performance in the infrared thermal radiation band and high transparency in the ultra-broadband longwave range (Fig. 1C). This capability is especially advantageous for widespread communication bands, including 5G (2.5 GHz - 39 GHz in the United States), GPS (~ 1.1 GHz - 1.5 GHz), Bluetooth (~2.4 GHz), and wireless LAN (2.4 GHz and 5 GHz).

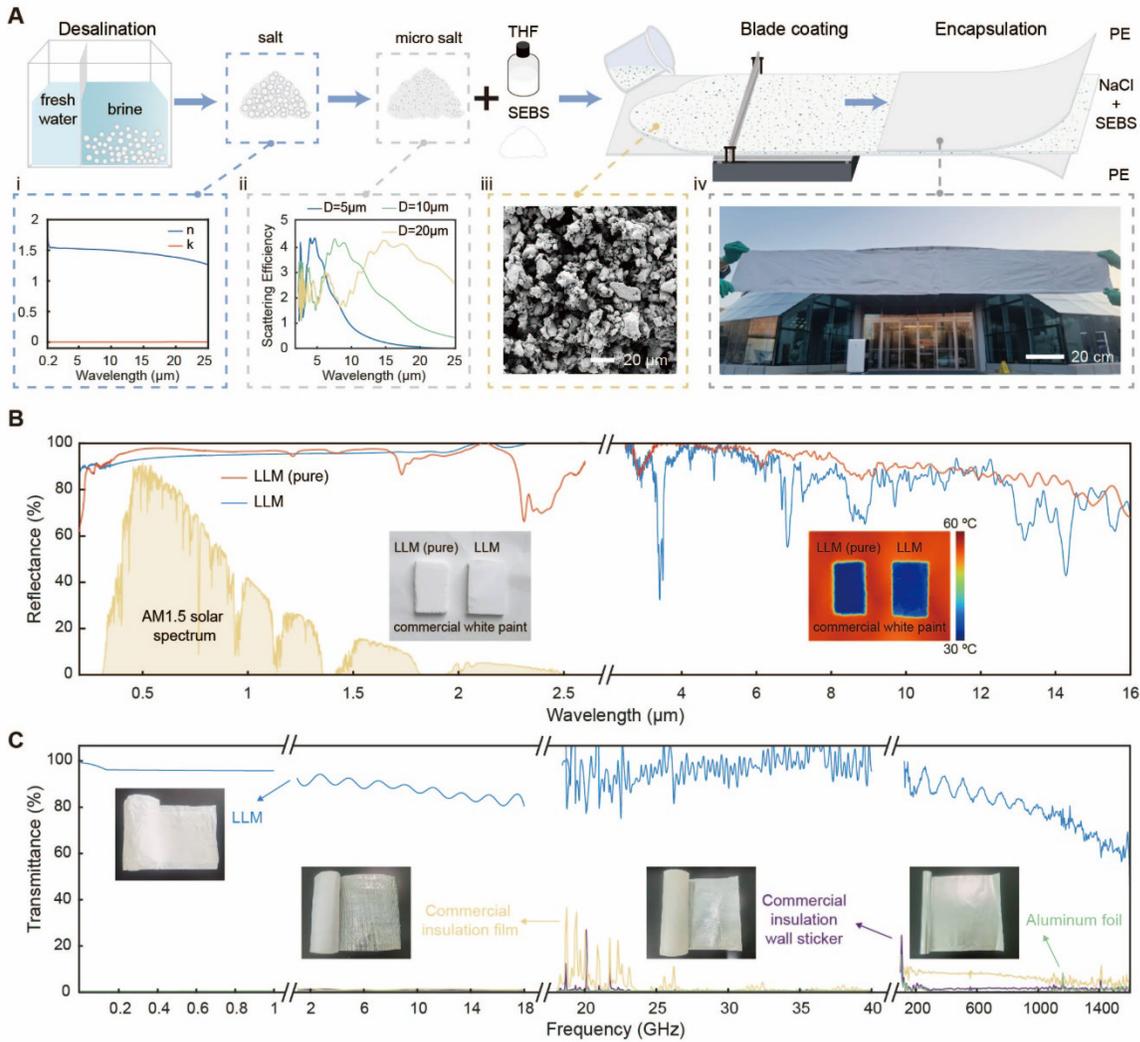

**Fig. 2 Fabrication and characterization of LLM.** (**A**) Schematic of the ultra-scalable blade coating method used to fabricate LLM in film form. The process employs NaCl microparticles sourced from desalination brine waste and uses nanoporous polyethylene (NanoPE) film as the packaging material. (i) The complex spectral refractive index ($n + ik$) of NaCl indicates negligible absorption across the solar to MIR spectrum (0.3 - 25 μm). (ii) Simulated scattering efficiency of circular NaCl microparticles with varying diameters across the 0.3 - 25 μm wavelength range. (iii) Scanning electron microscopy image showing the microstructure of NaCl microparticles embedded in a polymer binder (SEBS) within the LLM. (iv) Photograph of an LLM film measuring 2 m × 0.2 m. (**B**) Spectral reflectance of LLM across the solar to MIR spectrum (0.3 - 16 μm). Insets show (left) a photograph and (left) a thermography image of pure LLM sample and LLM film sample with high emissivity white paint as the background. (**C**) Spectral transmittance of LLM film across the longwave spectrum (100 kHz – 1.600 THz), compared with pure aluminum foil, commercial insulation film, commercial insulation wall sticker.

## Thermal insulation performance

To investigate thermal insulation performance of the LLM, we tested it under three scenarios (Fig. 3A). First, we assessed the LLM's ability to reduce heat loss in a customized metal box enclosure, simulating a typical room in a cold environment. The enclosure's inner surfaces were lined with insulation materials, with the low-e side facing inward to maximize radiative thermal insulation. A feedback-controlled heating system was set to 20 °C, and electricity usage was monitored continuously over a 48-hour period (Fig. 3B, fig. S13 and Supplementary Text 4). Compared to commercial white paint, the LLM reduced energy consumption by 41.1% over 48 hours, highlighting its superior radiative insulation capabilities (Fig. 3C). The LLM film also showed significant savings over commercial low-e insulation film, using 10.2% less electricity over the same period.

Next, we demonstrated the LLM's effectiveness in reducing heat gain in a hot environment. Enclosures were lined with insulation on both inner and outer surfaces, using low-e materials to mitigate heat gain from solar irradiation and the high-emissivity inner walls. To simulate minimal heat gain, such as low-temperature storage in cold chain logistics, we placed a 110 g ice cube inside the enclosure and monitored its mass over 7 hours (Fig. 3D and fig. S14). At the end of the test, the remaining mass of the ice cube in the LLM enclosure was 20.1 g, representing 64.8% and 52.3% higher retention than in enclosures with commercial white paint and state-of-the-art low-e insulation materials, respectively. This result demonstrates that the LLM's substantial cooling energy and cold-preserving potential, crucial for low-temperature storage applications (Fig. 3E and Fig. 3F). Additionally, we placed enclosures without active heating or cooling outdoors to assess temperature stability. The LLM enclosure displayed significantly smaller temperature fluctuations compared to commercial products, reducing fluctuations by 64% and 40%, respectively (Fig. 3G and Fig. 3H). This significant suppression of temperature fluctuations by the LLM enhances the energy efficiency of HVAC systems(*51*). We developed a thermal model (fig. S16 and Supplementary Text 5) to quantify potential HVAC energy savings when using LLM as inner and outer wall insulation for buildings under various climate conditions. The results indicate that combining the LLM envelope with low-e glass windows in buildings can achieve up to 13.9% of energy savings for space thermal management compared to a control building without low-e insulation.

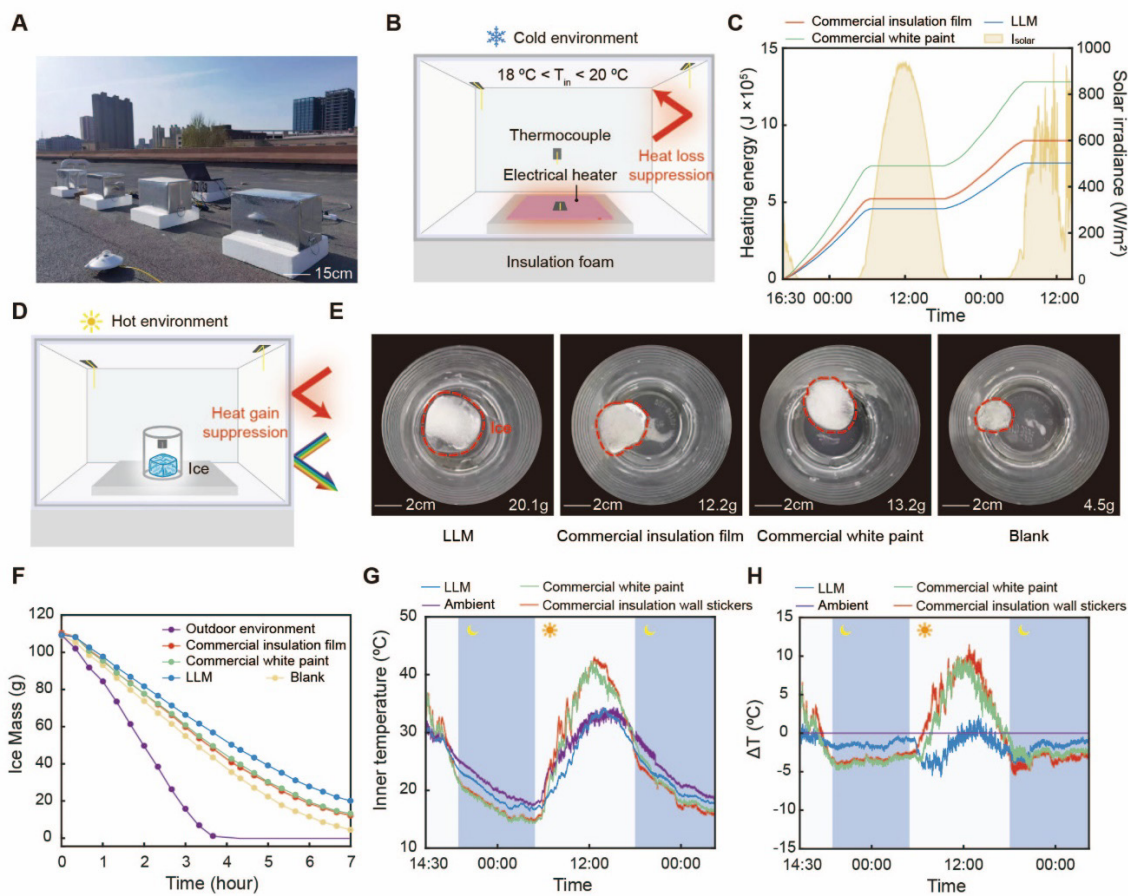

**Fig. 3. Thermal insulation performance of LLM in various scenarios.** (**A**) Photograph and (**B**) schematic illustration of thermal measurement system used to evaluate outdoor thermal insulation performance. Insulation materials are applied to the inner surfaces of building simulant boxes. A PID-controlled heating system maintains a constant temperature (20 °C) within the simulant boxes under cold outdoor conditions. (**C**) Cumulative heating power consumption over a 48-hour outdoor heat loss test. The LLM film reduces electrical heating power consumption by 41.1% compared to commercial white paint and 10.2% compared to commercial insulation film. (**D**) Schematic illustration of the heat gain test setup for the cold storage simulant box, containing ice cubes, under summer outdoor conditions. Insulation materials are applied to both the inner and outer surfaces of the box. (**E**) Photographs showing the melting of ice cubes in simulant boxes with different insulation materials after a 7-hour outdoor test. The ice cube in the box with the LLM exhibits minimal mass loss. (**F**) Measured mass change of ice cubes in simulant boxes with different insulation materials during the heat gain test. (**G**) Temperature variation of the enclosure over two days, including daytime and nighttime outdoor condition, with different insulation materials. (**H**) Temperature stability of the enclosure during the testing period. The LLM film reduces temperature fluctuations by 64% and 40% compared to commercial products, demonstrating its effectiveness in improving thermal stability by minimizing radiative heat gain and loss. Note that subambient temperature during the daytime are due to relatively low solar intensity, as shown in fig. S15.

**Versatile longwave management with LLM**

Beyond its exceptional thermal insulation performance, we demonstrated LLM's compatibility with longwave signal and power transfer, including applications in wireless communication, wireless power transfer, information identification and security screening. In the first scenario, we assessed the impact of low-e thermal insulation on wireless data transfer through building walls, utilizing a 26 GHz millimeter wave (MM-Wave) system to simulate 5G data transmission (Fig. 4A)(*52*). The experimental setup included a transmitter and a receiver with customized software to emit, receive and decode multimedia data. Two acrylic plates (1 m × 1 m), one with LLM and the other with a commercial low-e insulation wall sticker, were inserted manually between the transmitter and receiver (Fig. 4B, Supplementary Movie S1) during the transmission of a video of a rolling football at 357 kbps (Fig. 4C upper row, fig. S17). When the LLM was used as a wall barrier, the receiver captured the wireless signal successfully, allowing uninterrupted video playback (Fig. 4C left column). In contrast, the commercial low-e insulation wall sticker blocked the signal, causing the video to freeze and leaving the football's position unchanged (Fig. 4C bottom right). The constellation diagram for MM-Wave transmission showed that the LLM wall performed similarly to an ideal air scenario, with received symbol precisely matching their constellation points (Fig. 4D, fig. S18 and Supplementary Text 6). This is significantly contrast to the commercial insulation, where transmission was impaired by noise and interference, causing symbols deviation. These results highlight the superior longwave transmittance of the LLM, enabling high-quality wireless communication compared to state-of-the-art metallic low-e insulations.

LLM's high microwave transparency (Fig. 2C) enables applications in wireless microwave heating and power transfer. Unlike aluminum foil pans or insulation packaging, LLM does not interfere with microwave heating (Fig. 4E). To evaluate its effectiveness, a piece of sausage was reheated in a standard microwave oven (2.45 GHz) for 30 seconds using two containers: a paper bowl and an LLM bowl. Thermal images showed that the sausage in the paper bowl heated from 20.1°C to 65.1°C, a net increase of 45°C (Fig. 4E). In contrast, the sausage in the LLM bowl reached 69.6°C, indicating improved heating efficiency. Notably, the LLM surface remained relatively cool during the process, attributed to its low microwave absorption properties (Fig. 2C and fig. S19). This highlights LLM's potential to enhance microwave heating efficiency while protecting packaging materials from thermal damage. To further investigate the effect of different bowl materials on simultaneous microwave heating and insulation performance, 80 ml of water was heated in paper cups placed inside both an LLM bowl and a paper bowl, with temperature changes monitored in real time (Fig. 4F). After 1 minute heating in the microwave, the water in the LLM bowl reached 82.6°C, 1.9°C higher than the water in the paper bowl (80.7°C). After 30 minutes at room temperature, the water in the LLM bowl retained a temperature of 46.8°C, 5.7°C higher than the water in the paper bowl (41.1°C). These results confirm that LLM not only improves microwave heating efficiency but also offers superior thermal insulation, making it a promising material for applications requiring both efficient heating and sustained temperature retention.

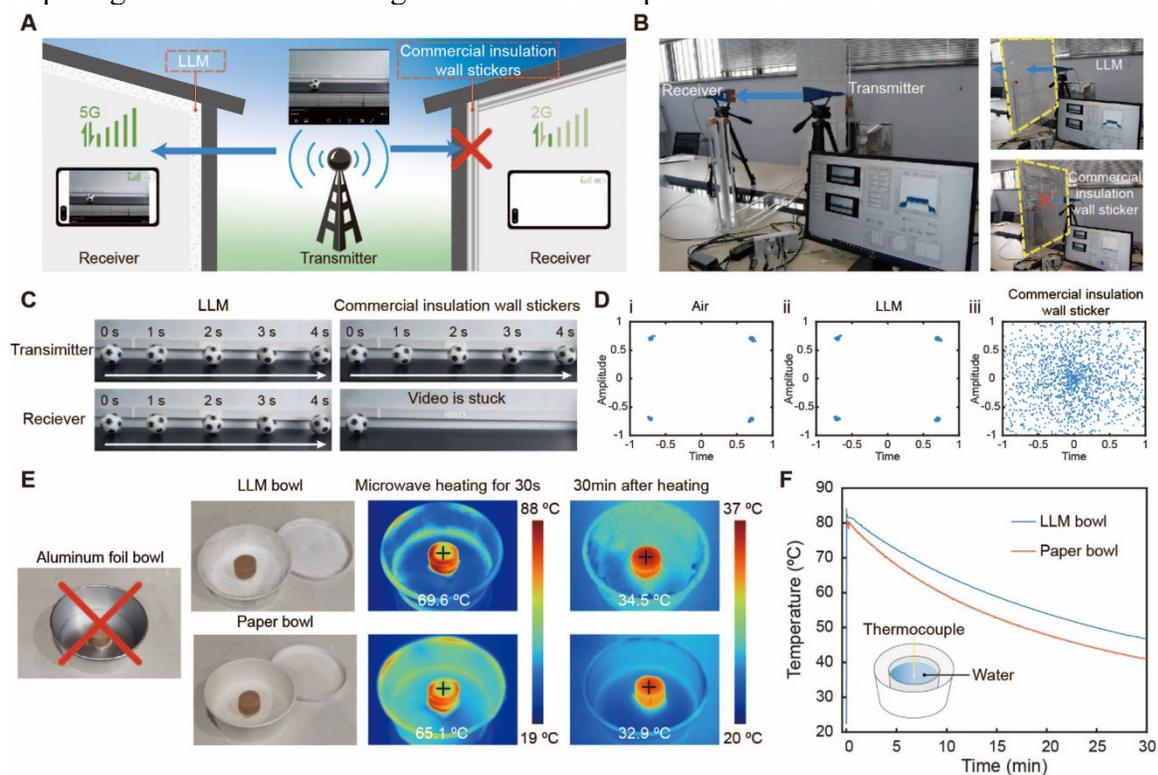

**Fig. 4. Demonstration of wireless communication and energy transfer with LLM.** (**A**) Schematic of real-time wireless communication for video data transfer between an outdoor transmitter and indoor mobile phones, through building walls equipped with either LLM or commercial low-e insulations. (**B**) Photographs of the wireless communication experiment setup. A millimeter wave (MM-Wave) system operating at 26 GHz simulates 5G data transfer. Both LLM and commercial low-e insulation materials are attached to 1 m × 1 m panel to model a building wall. (**C**) Time-series video data originally emitted by the MM-Wave transmitter (upper row) and received by the MM-Wave receiver (lower row). The video data passes through the LLM wall without interruption, but is blocked by the commercial low-e insulation wall. (**D**) Constellation diagrams of MM-Wave video transmission across air, LLM, and commercial low-e insulation walls. The LLM wall performs similarly to an air scenario, with minimal noise or interference, ensuring high-quality wireless communication. (**E**) Visual and thermographic images demonstrating microwave heating with different containers. The LLM-insulated bowl allows efficient microwave heating without interference, unlike aluminum foil bowls, which are unsuitable for microwave use. The sausage heated in the LLM bowl reaches a higher temperature,

and the bowl's surface remains relatively cool due to LLM's low microwave absorption properties. (**F**) Measured temperature variation of food and water in LLM and paper bowls before, during, and after microwave heating. The LLM bowl not only enhances heating efficiency – evidenced by higher temperatures during microwave heating – but also provides superior thermal insulation, maintaining higher temperatures over time compared to the paper bowl.

As a third practical application, we demonstrated the applicability of LLM in cold chain logistics, which involves managing temperature-sensitive products such as fresh food, chemicals, and pharmaceuticals. Due to the time-sensitive nature of modern cold chain logistics, rapid and non-invasive item tracking, identification, and inspection are crucial(*53, 54*). We tested the effectiveness of LLM for RFID and nondestructive terahertz security screening of items packed in insulation bags made with LLM and with aluminum foil. Temperature changes inside the bags were monitored to assess their ability to prevent heat transfer. Results showed that the LLM insulation bag maintained an internal temperature approximately 2°C lower than that of a standard commercial insulation bag over a 135-minute period at 40°C and 10% humidity, highlighting the LLM bag's superior ability to minimize heat gain (Fig. 5A). We next used an RFID system operating at 920 MHz to examine the effects of aluminum and LLM layers on RFID readability (Fig. 5B and fig. S20). The system included a handheld active reader and passive tags that optimized for distance responsiveness. The aluminum insulation bags completely shielded RF signals, resulting in a near-zero read range (Fig. 5C) due to the opacity of metals to longwave signals. In contrast, the LLM insulation bag had minimal impact on signal propagation, maintaining the same optimal reading range as the unshielded condition.

To further demonstrate the applicability of LLM in cold chain logistics, security screening and hazardous materials inspection, we employed a nondestructive THz imaging system to evaluate insulation materials' effect on imaging quality (Fig. 5D). Many materials have unique spectral fingerprints in the THz range, allowing precise material characterization and high-resolution imaging of object interiors, making this technology ideal for security screening applications(*55, 56*). We used a 0.035 THz imaging system to inspect items packed in insulation bags, including a foam board with various material pieces, a metallic toy gun, and a combination of a metallic toy gun, an orange, and a coke can (Fig. 5D and fig. S21). Results showed that the LLM insulation bag allowed clear imaging of metallic items, such as metal samples (Ag, Al, Cu), the metallic toy gun, and the coke can, due to metals' high reflectance in THz range (Fig. 5E). In contrast, the commercial aluminum foil bag completely blocked the THz signal, making content inspection unfeasible. Thus, using LLM in cold chain packaging enables effective security screening without requiring the bag to be opened, supporting the need for real-time inspection and thermal insulation in cold chain logistics.

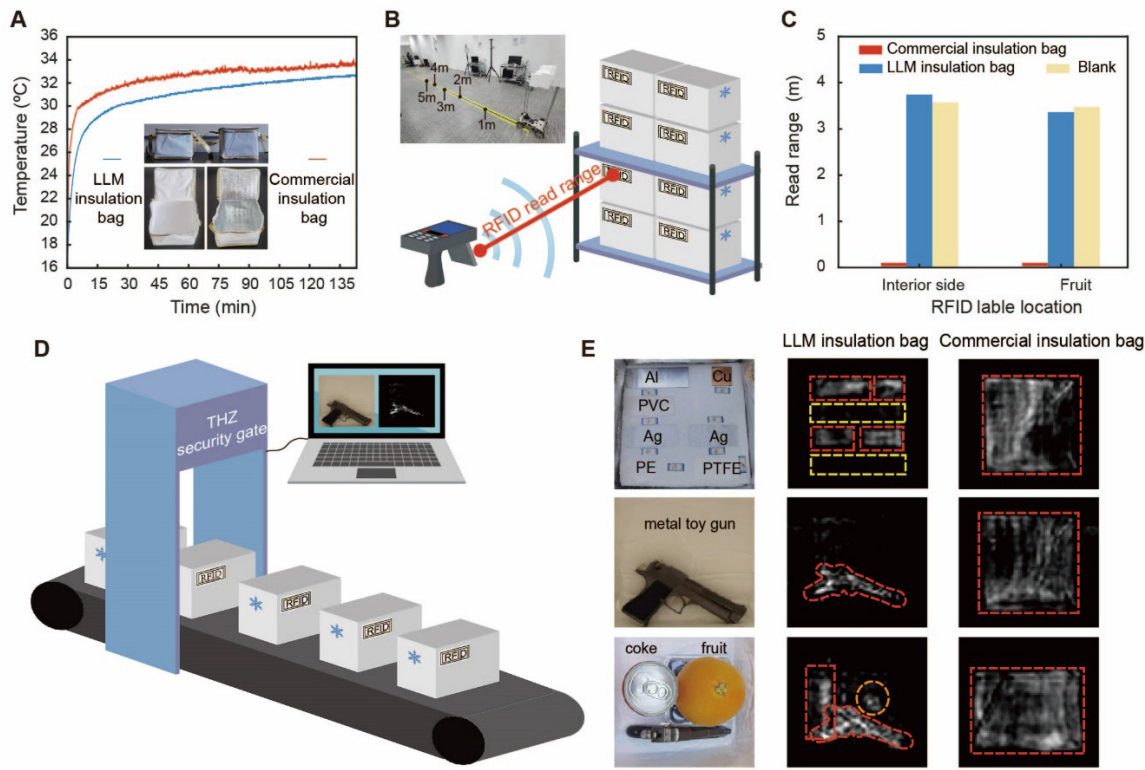

**Fig. 5. Wireless tracking and security screening in cold chain logistics with LLM.** (**A**) Temperature variation inside insulation bags in a high-temperature environment (40°C). The LLM insulation bag maintains a lower internal temperature (Approximately 2°C cooler) and rises more slowly compared to the commercial insulation bag, ensuring a stable low-temperature environment for cold chain products. (**B**) Schematic of an RFID system for automated tracking and identification of tags attached to storage boxes. The inset photograph shows the experimental setup for RFID tag reading with sample insulation bags. (**C**) Comparison of RFID signal read range for various insulation materials and RFID tag attachment positions. (**D**) Schematic of nondestructive THz imaging for security screening of potential hazardous items in cold chain logistics, such as explosives, firearms, drugs and biochemical weapons. (**E**) Photographs of raw metal materials (aluminum, silver, copper, top left), a prohibited metallic toy gun (middle left) and mixed items (orange, coke can, metallic toy gun; bottom left) placed on a THz imaging platform operating at 0.035 THz. Inspection results: items in the LLM-insulated bags (middle column) display clear morphological details, while those in aluminum foil bags (right column) are completely obscured by the aluminum layer.

## Discussion

The presented LLM introduces a novel photonic concept that achieves low emissivity through all-dielectric thermal photonic structures while maintaining ultra-broadband longwave transparency. Building upon this foundational work, we envision a wide range of new LLMs can be readily deployed in various forms and functionalities, by leveraging recent advances in thermal photonics and radiative cooling over the past decade(*4, 5*), including photonic designs utilizing strong scattering-based reflection(*7-9, 12, 13*), materials (e.g. KBr based LLM, fig. S22)(*20, 30, 36, 57, 58*) and scalable manufacturing techniques(*7, 10, 11, 59, 60*). This approach has the potential to significantly impact interdisciplinary fields such as photonics, thermal, information and material science. More broadly, the capabilities of LLM contribute to carbon neutrality, smart city development and enhanced digital connectivity by revolutionizing applications in building insulation, electrified transportation, cold chain logistics, wireless charging (fig. S23 and Supplementary Movie S2), wireless doorbells (Supplementary Movie S3 and Supplementary Text 7), and even camouflage and stealth technologies.

## Materials and Methods

### Fabrication of the LLM

The NaCl microparticles were prepared by physically crushing raw salt crystals (99.99%, Aladdin) followed by a screening process using a fine wire mesh to achieve particle sizes in the range of 700 nm to 40 μm (see fig. S1 and fig. S2 for images of NaCl microparticles and their size distribution).

To fabricate pure LLM, the prepared NaCl microparticles were hydraulically compressed into the desired shape for the bulk precursor to sintering. The precursor was placed into a tubular furnace (BTF-1200C, AnHui BEQ Equipment Technology Co., Ltd) for sintering, where the temperature was steadily raised from room temperature to 500 °C, held for 3 hours, and then cooled back to room temperature.

To fabricate scalable LLM, styrene-ethylene/butylene-styrene (SEBS) polymer (Kraton, G1650MU) (fig. S5) was completely dissolved in a solvent (i.e., THF or acetone) under magnetic stirring at room temperature for 12 hours. The prepared NaCl microparticles were then added to the solution and stirred until fully dispersed. The precursor mixture was blade coated onto a nanoporous polyethylene (NanoPE) film (San Yuan material company, thickness: 25 μm) (fig. S6) fixed on a glass/PTFE plate. After coating, another piece of NanoPE film was placed on top for encapsulation. The solidified film form of LLM was obtained after the solvent completely evaporated at room temperature over 12 hours.

### Material characterization

Reflectance and transmittance spectra in the UV-Vis-NIR range (200 nm to 2500 nm) were measured using a spectrophotometer (Cary 5000 UV-Vis-NIR, Agilent) equipped with an integrating sphere (Internal DRA2500, Agilent). Infrared reflectance and transmittance spectra were obtained using an FTIR spectrometer (VERTEX 80V, Bruker) with a gold integrating sphere (A562, Bruker). Thermographic images were captured using an infrared camera (Ti480 PRO, Fluke) operating in the wavelength range of 7.5-14 μm. The microstructure images of LLM were taken using scanning electron microscopy (AURIGA-4506, ZEISS). The water contact angle was determined with a goniometer (JC2000D3, Powereach). The size distribution of NaCl crystals was analyzed by a laser particle sizer (Mastersizer 3000, Malvern).

### Selection of control group samples

We used commercially available radiant barriers and a commercial solar reflective white paint as control groups in our thermal insulation and wireless communication experiments. As shown in fig. S10, the commercially available radiant barriers included pure aluminum foil, commercial insulation film (Peng Yu Bag Factory, 2 mm thick), and reflective commercial insulation wall stickers (Hexin Home Furnishings Co., Ltd., 2 mm thick). These were used as control groups in Fig. 2C, Fig. 3A – 3H, and Fig. 4B – 4D in the main text, as well as in fig. S10-S14, fig. S17, and fig. S18. The commercial white paint (DutchBoy Maxbond Exterior Flat White Paint) was used as a control group in Fig. 2B and Fig. 3A – 3H. In the energy transfer performance experiment, we used paper bowls and aluminum foil bowls (Huamei Haoli Commercial Center) as control groups in Fig. 4E and Fig. 4F. To demonstrate the wireless tracking and security inspection capabilities, we used a commercial insulation bag (Jingdong Haiyuan Trading Co., Ltd.) as a control group in Fig. 5A – 5E.

### Characterization of transmission spectra for radio waves

We tested the electromagnetic shielding effectiveness (S factor) in the 300 KHz-1 GHz range using a coaxial method with a microwave network vector analyzer (Agilent, E5071E). The S parameters obtained from the test were converted into transmission spectra.

The EM transmittance spectra of LLM films and control groups were analyzed using a free space method, measuring at normal incidence. Two lens antennas with frequency ranges of 1-18 GHz and 18-40 GHz were used to accommodate the bandwidth limitations of a single antenna. These lens antennas were connected to a vector network analyzer (N5230C) via a phase-stable cable, which transmitted and received a continuous swept-frequency signal. For calibration and normalization of reflectivity, two measurements were performed: one with the sample and one with a metal plate for comparison.

To investigate THz transmission spectra, we used a fiber-coupled terahertz time-domain spectrometer (University of Shanghai for Science and Technology) to test the transmission spectra of LLM and control groups in the 0.1-1.6 THz range, as shown in Fig. 2C. The THz transmittance of the LLM was significantly higher than that of the commercial metal low emissivity film. The terahertz time-domain spectrometer had a resolution of less than 5 GHz and a signal-to-noise ratio of greater than 90 dB.

**Outdoor thermal insulation test**

An outdoor thermal test of the thermal insulation performance of LLM films was conducted in Changchun, China. Fig. 3A shows the setup for the heat loss test. Insulation foam was placed between the building simulant box (25 cm width, 40 cm length and 24 cm height) and the ground to minimize conductive heat exchange. The inside surfaces of these boxes were covered with insulation films. A PID-controlled heating system, consisting of a power supply, computer, electric heater, thermocouple, and data logger (RDXL6SD, Omega), was used to maintain a constant temperature of 20°C in the building simulant box under cold outdoor conditions. Thermocouples were attached at identical positions in each box and the electric heater (heater size: 10 cm by 10 cm) to measure the internal and heating temperatures for parallel comparison. All thermocouples were precisely calibrated before measurements. A pyranometer (CMP 6, Kipp & Zonen) placed on the roof measured solar irradiance, with a data logger rated for a directional error of ±20 W/m².

Fig. 3D shows the setup for the heat gain test in a hot environment. A 110 g ice cube was prepared using a mold and placed in a cup. Both the inner and outer surfaces of four boxes were covered with insulation films. The mass of the ice was recorded every 20 minutes over a 7-hour period. A thermocouple was attached to the ground surface to record ground temperature, and another thermocouple was placed in the air to measure air temperature.

**Millimeter wave communication test**

The millimeter wave communication system used to simulate mobile communication data transfer consisted of two millimeter wave mixers, two horn antennas, and a software-defined radio (USRP-2974), as shown in Fig. 4A. The transmitter, located on the left side of the system, encoded and modulated the video signal to a 1.8 GHz carrier frequency using the host computer. This signal was then upconverted to 27 GHz by a mixer connected to one port of the USRP and transmitted into free space via the horn antenna. On the right side, the receiver captured the 27 GHz communication signal through the right horn antenna, down converted it to 1.8 GHz using a mixer, and input it into the two ports of the USRP, where it was sampled and demodulated.

The communication system employed QPSK (Quadrature Phase Shift Keying) modulation technology. The QPSK constellation consisted of four points, each representing a different phase state. Each dot in the constellation diagram represented a symbol, with each symbol carrying two bits of information. When the communication link was obstructed, the transmitted signal energy could not reach the receiver, resulting in the received signal energy being lower than the noise level. This caused the constellation diagram to become disordered, as shown in Fig. 4D(iii). In this situation, the receiver could not correctly demodulate the bit information transmitted by the transmitter, rendering the communication system non-functional. Conversely, when the communication link was unobstructed, the transmitted signal energy reached the receiver, resulting in a high signal-to-noise ratio. The points in the constellation diagram appeared scattered and stable, as shown in Fig. 4D(ii). In this scenario, the receiver could correctly demodulate the bit information transmitted by the transmitter, allowing the communication system to function normally.

**Microwave heating demonstration**

The LLM bowl was prepared by cutting off the main part of a paper bowl, leaving only a one-centimeter-wide skeleton, and attaching the LLM film to this framework. A 50 g slice of sausage was placed at the bottom of each bowl, covered, and heated in a microwave for 30 seconds. After heating, the bowls containing the sausage were placed at room temperature. Infrared thermal images of the sausages and bowls before and after heating were taken using an infrared camera (Ti480 PRO, Fluke).

Two cups containing 80ml of water was placed into paper bowls and LLM bowls respectively. After covering the bowls with lids, they heated in a microwave for one minute, and then were placed at room temperature. Thermocouples were positioned in the water to record the temperature.

**Insulation bag thermal performance test**

The outer surface of the commercial thermal insulation bag was made of non-woven fabric, and the inner surface was lined with a commercial thermal insulation film, which included a reflective layer of aluminum foil. Similarly, the outer surface of the LLM insulation bag was also made of non-woven fabric, but its inner surface was lined with LLM film. As shown in Supplementary Fig. 20A, items such as a Cola can, blueberries, and grapefruit were placed in both insulation bags. After thermal equilibrium was reached at room temperature, the bags were placed in an environmental chamber set to 40°C with 10% humidity. Thermocouples were inserted into the center of each bag to measure the internal temperature for parallel comparison.

**RFID read range test**

RFID tags (Wuxi GRAND-TAG Electronic Technology Co., Ltd.) were attached to both the inside and outside surfaces of the insulation bag, as shown in Supplementary Fig. 20A. A handheld reader (CHAINWAY, C72) was used to read the radio frequency signals for the read range test. The Tagformance system was set to sweep from 800-1000 MHz, and the C72 operated at 920 MHz with a power of 30 dBm. Theoretical forward and reverse read ranges of the RFID tags were measured using the Tagformance Pro system (Voyantic).

**THz security imaging test**

fig. S21 illustrates the THz security imaging system. Commercial and LLM insulation bags containing various items were placed into the system for security imaging using terahertz radiation. The items tested included metal toy guns, fruits (blueberries and grapefruits), a Cola can, and prepared test plates. The test plates were created by attaching square films of different materials—such as silver foil, aluminum plate, copper plate, PVC, PTFE, and PE—to various positions on a foam plate. Each item was placed individually or in combination at the bottom of the commercial and LLM insulation bags for THz security imaging.

## Acknowledgments

### Funding

National Natural Science Foundation of China grant T2525033
National Natural Science Foundation of China grant 62134009
National Natural Science Foundation of China grant 62121005
National Natural Science Foundation of China grant 62475261
National Natural Science Foundation of China grant 62288101
International Partnership Program of Chinese Academy of Sciences or Future Network Grant No. 171GJHZ2023038FN

### Author contributions

Conceptualization: W.L., T.C., L.L.
Methodology: Y.Z., L.L., J.D., X.Z., Q.Z., N.Y., R.J., J.W., X.L., M.S., X.K., Y.L., Q.C., Y.Z., T.C., W.L.
Investigation: Y.Z., L.L., J.D., X.Z., Q.Z., N.Y., R.J., X.L., M.S., J.W., X.L., X.K., Z.L., Y.L., Q.C., Y.Z., T.C., W.L.
Visualization: Y.Z., L.L., J.D., F.Z., J.W., X.L., X.K., W.L.
Funding acquisition: W.L., L.L., T.C.
Project administration: W.L.
Supervision: W.L., T.C., L.L.
Writing – original draft: W.L., L.L., Y.Z.
Writing – review & editing: W.L., T.C., L.L., J.D., Y.Z.

### Competing interests

The authors declare that they have no competing interests.

### Data and materials availability

All data are available in the main text or the supplementary materials. Information requests should be directed to the corresponding authors.

## Supplementary Materials

Supplementary Text
Figs. S1 to S23
Movies S1 to S3